\documentclass[a4paper]{jpconf}
\usepackage{graphicx}

\begin{document}


\title{Geometrical clusterization of  Polyakov loops in SU(2)  lattice gluodynamics}
\author{A Ivanytskyi$^{1, 2}$, K Bugaev$^1$, E Nikonov$^{3}$, E-M Ilgenfritz$^{4}$,
            V Sagun$^{1, 5}$, I Mishustin$^{6}$, V Petrov$^{1}$, G Zinovjev$^{1}$}
\address{$^1$ Bogolyubov Institute for Theoretical Physics, Metrologichna str. 14b, Kyiv, 03680, Ukraine}
\address{$^2$ Astronomical Observatory of Taras Shevchenko National University of Kyiv, Observatorna 3, Kyiv, 04053, Ukraine}
\address{$^3$ Join Institute for Nuclear Researches, LIT, Dubna, 141980, Russia}
\address{$^4$ Join Institute for Nuclear Researches, BLTP, Dubna, 141980, Russia}
\address{$^5$ Centro Multidisciplinar de Astrof{\'i}sica, Instituto Superior T$\acute{e}$cnico, Universidade de Lisboa,
Av. Rovisco Pais 1, 1049-001 Lisboa, Portugal}
\address{$^6$ FIAS, Ruth-Moufang-Strasse 1, Frankfurt upon Main, 60438, Germany}
\ead{aivanytskyi@bitp.kiev.ua, bugaev@fias.uni-frankfurt.de}


\vspace*{-.2cm}

\begin{abstract}
The liquid droplet formula is applied to an analysis of the properties of geometrical 
(anti)clusters formed in SU(2) gluodynamics by the Polyakov loops of the same sign. 
Using this approach, we explain the phase transition in SU(2) gluodynamics as a 
transition between two liquids during which one of the liquid droplets (the largest 
cluster of a certain Polyakov loop sign) experiences a condensation, while the droplet of another 
liquid (the next to the largest cluster of the opposite sign of Polyakov loop) evaporates. 
The clusters of smaller sizes form two accompanying gases, which behave oppositely 
to their liquids. The liquid droplet formula is used to analyze the size distributions of
the gaseous  (anti)clusters. The fit of these distributions allows us to extract the temperature 
dependence of surface tension and the value of Fisher topological exponent $\tau$ for 
both kinds of gaseous clusters. It is shown that the surface tension coefficient of 
gaseous (anti)clusters can serve as an order parameter of the deconfinement phase 
transition in SU(2) gluodynamics. The Fisher topological exponent $\tau$ of clusters 
and anticlusters 
is found to have the same value 1.806 $ \pm$  0.008. This value disagrees with the famous 
Fisher droplet model, but it agrees well with an exactly solvable model of the nuclear 
liquid-gas phase transition. This finding may evidence for the fact that the SU(2) 
gluodynamics and this exactly solvable model of nuclear liquid-gas phase transition 
are in the same universality class.
\end{abstract}

\vspace*{-.8cm}

\section{Introduction}

The lattice simulations are presently  considered as the only first principle tool to investigate the deconfinement phase
transition (PT) in quantum chromodynamics (QCD). Such a PT is also expected in gluodynamics (GD) which is a pure non-Abelian gauge theory. The Svetitsky-Jaffe hypothesis \cite{Yaffe82a,Yaffe82b} relates the deconfinement PT in  
(d+1)-dimensional SU(N) GD to the magnetic PT in Z(N) symmetric spin model in d-dimensions. 
The local Polyakov loops  in  GD are playing  the 
role of spins in  Z(N) symmetric spin model.
For the (d+1)-dimensional  lattice having the size $N_\sigma^d\times N_\tau$ the local Polyakov loop is defined by the temporal gauge links $U_4(\vec x,t)$ as\vspace*{-.15cm}
\begin{equation}
L(\vec x)=Tr\prod_{t=0}^{N_\tau-1} U_4(\vec x, t). \vspace*{-.2cm}
\end{equation}
A high level of understanding of the spin systems along with the Svetitsky-Jaffe hypothesis led  to a significant 
progress in studying  the SU(N) GD properties in the PT vicinity. Formation of geometrical clusters composed 
of the Polyakov loops is an important feature of GD \cite{Fortunato00,Fortunato01}. A similar phenomenon is 
well known in spin systems and it is responsible for percolation of clusters, which already was studied in GD
with paying a special attention  to the largest and the next to the largest clusters \cite{Gattringer10,Gattringer11}. 
However,  already from  the  famous  Fisher Droplet Model (FDM) \cite{Fisher67,Fisher69}
it is  well-known that some essential features of  the liquid-gas PT
are encoded in  the properties of smaller clusters. 
A principally important feature   of the FDM  (and all its followers)  is that 
at  the critical point the  size distribution of physical clusters obeys a power law which is controlled by the Fisher 
topological exponent $\tau$. Hence the value  of $\tau$ maybe a key element  both  for  developing  a consistent 
cluster model of  liquid-gas PT  for  QCD and for experimental  localization of  the QCD critical point with its help. Therefore, in this work  we study the geometrical clusterization in SU(2) GD and analyze the properties of  clusters of all possible sizes. This approach allows us to explain the deconfinement of color charges as   the liquid-gas PT \cite{Ivanytskyi} of special kind. \vspace*{-.4cm}

\section{The Polyakov loop geometrical clusters}

In case of SU(2) gauge group $L(\vec x)$ has the  real values  from  $-1$ to $1$. Similarly to spin models, for a given lattice 
configuration the  Polyakov loops being the nearest neighbors  can be attributed to the same  cluster, if they have the same sign. The boundaries of clusters with opposite signs of $L(\vec x)$ are characterized by  strong fluctuations. Therefore, 
similarly to Refs. \cite{Gattringer10,Gattringer11} we introduced the minimal absolute value of the Polyakov 
loop attributed to the clusters, i.e. a cut-off $L_{cut} > 0$. All space points $\vec x$ with $|L(\vec x)| \le L_{cut}$ 
are attributed  to ``auxiliary'' or ``confining'' vacuum which volume fraction is independent of the inverse lattice coupling 
$\beta=\frac{4}{g^2}$ \cite{Ivanytskyi}, where $g^2$ is the lattice coupling constant. The above definition allows 
us to define the monomers, the dimers, etc. as the clusters made of a corresponding number of ``gauge spins" of the 
same sign  which are  surrounded  either by the  clusters of  opposite ``spin''  sign or  by a vacuum \cite{Ivanytskyi}.  
Obviously, there are  clusters  of two types related to two signs  of the local Polyakov loop. We introduce a formal 
definition of  anticlusters, if  their sign coincides with the sign of the largest n-mer existing at a given lattice configuration. 
The largest anticluster is  the ``anticluster droplet" and the  other n-mers (n =1, 2, 3,...) of the same sign correspond to the 
``gas of anticlusters". The  clusters are defined to  have an  opposite sign of the Polyakov loop and the largest of
 them is called  the ``cluster droplet". \vspace*{-.4cm}

\begin{figure}
\begin{center}
\includegraphics[width=35pc]{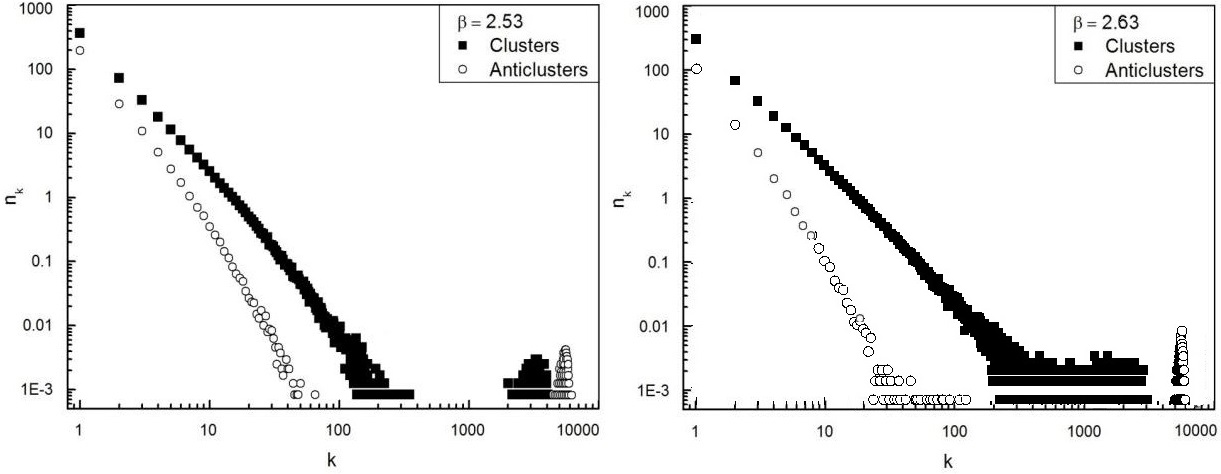}
\end{center}
\vspace*{-.8cm}
\caption{Typical size distributions of clusters and anticlusters for $\beta=2.53$  (left  panel) and $\beta=2.63$ 
(right  panel) are shown for the  cut-off $L_{cut}=0.2$.}
\vspace*{-.6cm}
\label{Ivan_Fig1}
\end{figure}

\section{Size distributions of clusters}

The described  scheme  of the (anti)cluster identification was realized numerically. The Polyakov loops were obtained 
at each spatial point of 3+1 dimensional lattice with the spatial and temporal extents   $N_\sigma=24$ and $N_\tau=8$,
respectively. The simulations were performed for 13 values of the inverse lattice coupling $\beta$ inside the interval 
$\beta\in[2.31,3]$. The physical temperature $T$ is  defined via two-loop $\beta$ dependence of 
the lattice 
spacing $a(\beta)$ as $1/T=N_\tau a (\beta)$ \cite{Gattringer10}.  The $\beta$ points were distributed not uniformly. They were concentrated in the 
PT region which is of principal interest for  this study. The identification of (anti)clusters was performed for two values 
of the Polyakov loop cut-off $L_{cut}=0.1$ and $L_{cut}=0.2$. For most of $\beta$ values the number of (anti)clusters of each  size was averaged over  the ensemble of 800 and 1600 gauge field lattice configurations. The  distributions  obtained in 
this way were the same within the statistical errors. 
For the gaseous anticlusters at three largest values of  $\beta$ the statistics was increased to 2400 
configurations. The right hand side vicinity of  PT was also analyzed with such a statistics in order to  exclude the effects 
of numerical fluctuations which we observed for $\beta=2.52,~2.53$ and $2.67$. The typical size distributions of
(anti)clusters for $L_{cut}=0.2$  are shown in Fig. \ref{Ivan_Fig1}. For $\beta$ below the critical value 
$\beta_c^\infty=2.5115$  in an  infinite system
\cite{Fingberg} the distributions of (anti)clusters are identical due to existing  global Z(2) symmetry. If $\beta$ is even 
slightly above $\beta_c^\infty$ then the symmetry between (anti)clusters breaks down (the left panel of Fig. \ref{Ivan_Fig1}). 
In this case the size of the cluster (anticluster) droplet decreases (increases). It is remarkable, that the corresponding gaseous clusters  behave contrary to their droplets. Therefore,  the deconfinement PT in SU(2) GD  can be 
considered as an  evaporation of the cluster droplet into the gas of clusters  and  a simultaneous condensation of the gas 
of  anticlusters  into the anticluster droplet. For large values of  $\beta$ (the right panel of Fig. \ref{Ivan_Fig1}) the cluster droplet becomes indistinguishable from its gas whereas the size of  anticluster droplet becomes comparable to the system size. 
Also  we found  that the gas and liquid  branches of anticluster distributions are always  well separated from each other. 
\begin{figure}
\begin{center}
\includegraphics[width=33pc]{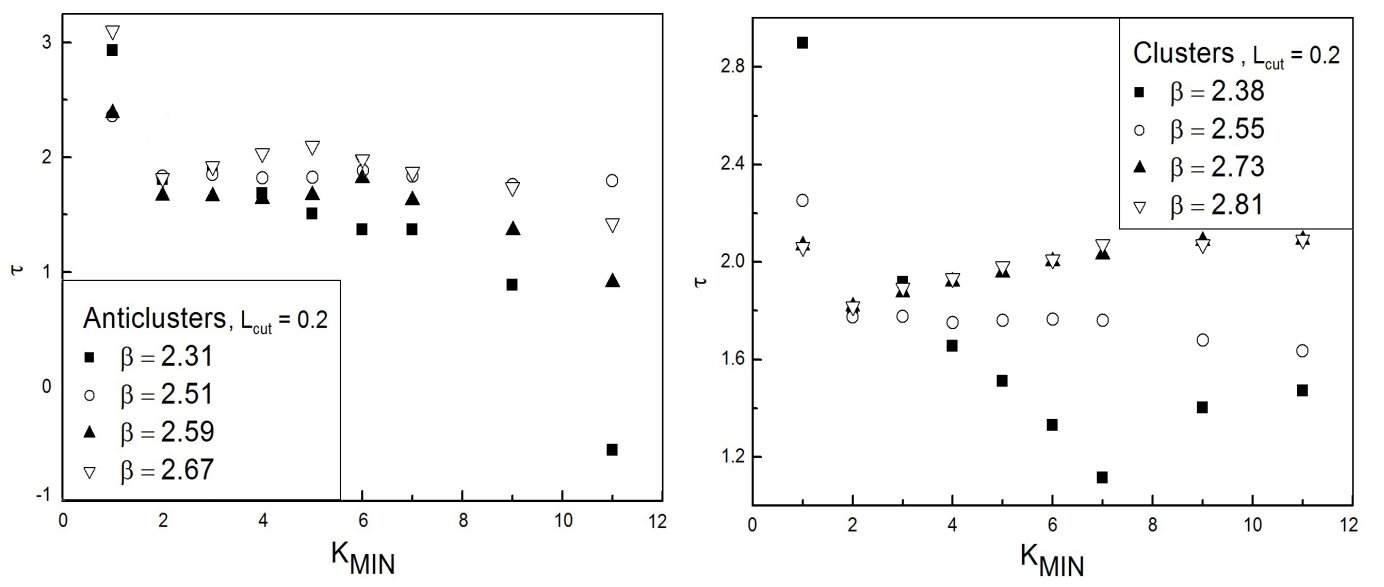}
\end{center}
\vspace*{-.8cm}
\caption{The Fisher exponent $\tau$ for several values of $k_{min}$ and for a few values of $\beta$ found by the 
4-parametric fit of the LDM formula. All these results refer to the cut-off $L_{cut}=0.2$.}
\vspace*{-.5cm}
\label{fig3}
\end{figure}

The found distributions closely resemble the ones discussed  for the nuclear fragments  and for the Ising  spin 
clusters \cite{Sagun}. Hence,  we analyzed whether the Liquid Droplet Formula (LDF) \cite{Fisher67} \vspace*{-.15cm}
\begin{equation}
\label{EqII}
n_A^{th}(k)=C_A\exp\left(\mu_A k-\sigma_A k^{2/3}-\tau \ln k\right) \,, \vspace*{-.2cm}
\end{equation}
is able to reproduce the size distributions  of clusters ($A=cl$) and anticlusters ($A=acl$). Here $T\, \mu_A k$ is the 
bulk part of free energy of k-mer, $T\, \sigma_A k^{2/3}$ denotes its surface free energy with the surface 
proportional to $k^{2/3}$, $\tau$ is the Fisher topological constant and $C_A$ is the normalization factor.
Since the LDM is valid only for (anti)clusters  which are  sufficiently large, then the question which has  to be clarified  
first  was a determination of the minimal size $k_{min}$ to  which Eq. (\ref{EqII}) can be applied. The standard procedure 
of $\chi^2/dof$ minimization with respect to $C_A$, $\mu_A$, $\sigma_A$, $\tau$ and $k_{min}$ allowed us to answer 
this question. Our analysis shows that the LDF is already able to describe the dimers. Indeed, we found that 
$\chi^2/dof\approx 1$  for all $k_{min}\ge2$ , whereas for $k_{min}=1$ we got $\chi^2/dof\approx 10$ which 
corresponds to the low quality of  data description. Within the statistical errors we found that $\tau$   is independent 
on $\beta$  for $k_{min}=2$ (see Fig. \ref{fig3}). This remarkable result is in line with  the predictions of cluster 
models \cite{Fisher67,Fisher69,Sagun}. Moreover,  we found  that $\tau<2$ both  for clusters and for anticlusters, 
which agrees  with the  exactly solvable model of the nuclear liquid-gas PT \cite{Sagun} and contradicts to the FDM 
\cite{Fisher67,Fisher69}. Thus, from a four parametric fit of the LDF we found that $k_{min}=2$ and $\tau=1.806(8)$ 
both for  clusters and for anticlusters. 

\begin{figure}
\begin{center}
\includegraphics[width=34pc]{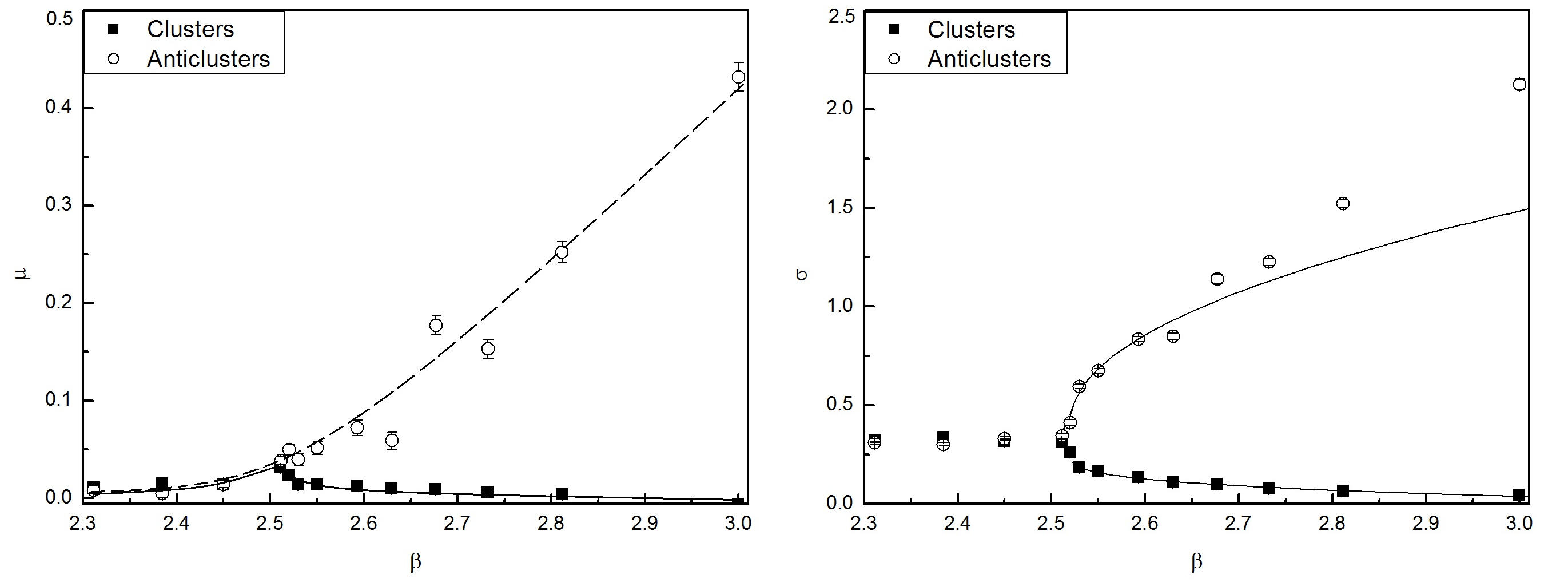}
\end{center}
\vspace*{-.8cm}
\caption{Reduced chemical potential (left panel) and reduced surface tension  (right panel) as function of
$\beta$ obtained for $L_{cut}=0.2$. The curves shown in the right panel represent Eq. (\ref{VI}).}
\vspace*{-.4cm}
\label{figfig}
\end{figure}

For fixed values $k_{min}=2$ and $\tau=1.806(8)$  we performed a three parametric fit 
of the (anti)cluster size distributions  to define $C_A$, $\mu_A$ and $\sigma_A$ with high precision. The typical value 
of $\chi^2/dof \approx 1$ was obtained for any  $\beta$, which signals  about high quality of the data description.
The $\beta$-dependences of $\mu_A$ and $\sigma_A$ are shown in Fig.  \ref{figfig} for $L_{cut}=0.2$. For $L_{cut}=0.1$
the results are similar. \vspace*{-.4cm}

\section{Order parameters of the deconfinement  phase transition}

\begin{figure}
\begin{center}
\includegraphics[width=17pc]{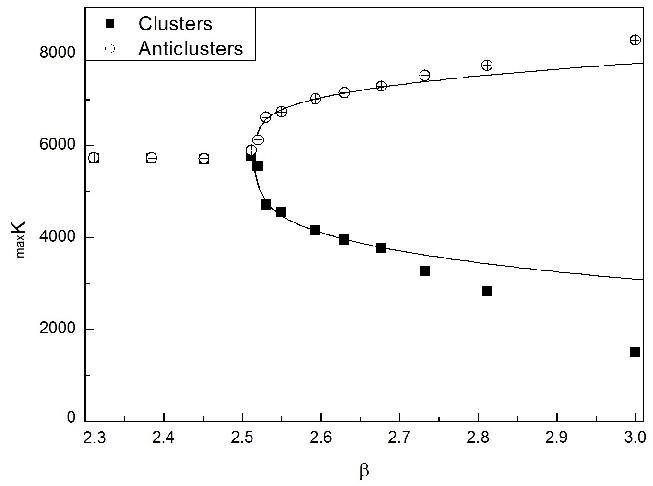}
\includegraphics[width=17pc]{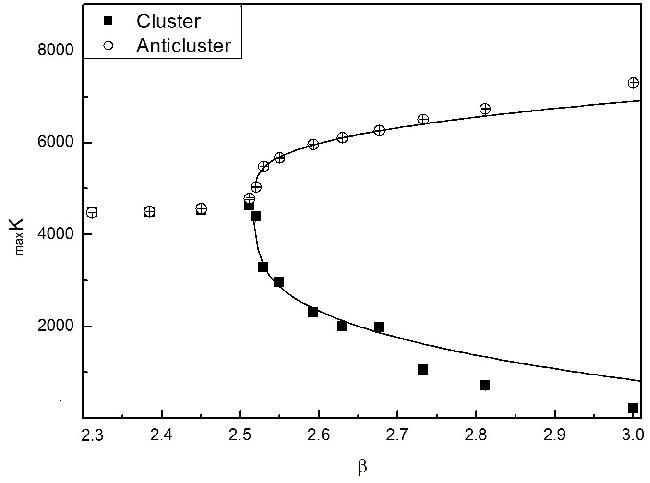}
\end{center}
\label{fig5}
\vspace*{-.75cm}
\caption{Dependence of the mean  size of the maximal (anti)cluster found for $L_{cut}=0.1$ (left panel) and for 
$L_{cut}=0.2$ (right panel). The curves represent Eq. (\ref{VII}).}
\vspace*{-.7cm}
\end{figure}

From Fig.  \ref{figfig} it is seen that the behavior of the reduced surface tension coefficient 
drastically changes, if  $\beta$ exceeds the critical value $\beta_c=2.52$. Indeed, for  $\beta\le\beta_c$ this 
quantity is constant and it is identical for clusters and anticlusters, while $\sigma_{cl}$ monotonically decreases and $\sigma_{acl}$ monotonically increases with $\beta$  for $\beta>\beta_c$. The qualitatively different  behavior of $\sigma_{cl}$ 
and $\sigma_{acl}$ allows us to treat them as an order parameter of the deconfinement PT in SU(2) GD. 
On the other hand, the average value of  Polyakov loop  $\langle L(\vec x) \rangle$ is traditionally considered as an order parameter of the deconfinement PT in SU(2) GD. In \cite{Ivanytskyi} we demonstrated  that  $\langle L(\vec x) \rangle$ 
is mainly defined by the largest (anti)cluster droplets whose  average size (see Fig. \ref{fig5}) is  given by 
\vspace*{-.2cm}
\begin{equation}
\label{EqV}
\max K_A=\sum\limits_{k=1} k^{1+\tau}n_A(k)\Biggl/\sum\limits_{k=1} k^{\tau}n_A(k) \,. \vspace*{-.3cm}
\end{equation}
The $\beta$ dependence  of $\sigma_A$ and $\max K_A$ in the right  hand side vicinity of $\beta_c$ is parametrized as
\vspace*{-.2cm}
\begin{eqnarray}
\label{VI}
\sigma_A(\beta)&=&\sigma_A(\beta_c)\pm d_A (\beta-\beta_c)^{B_A},\\
\label{VII}
\max K_A(\beta)&=&\max K_A(\beta_c)\pm a_A (\beta-\beta_c)^{b_A} \,
\end{eqnarray}
\vspace*{-.7cm}\\
where the signs ``+" and ``-" correspond to A=acl and A=cl, respectively,  and $b_A$ and $B_A$ are the critical exponents,
whereas $d_A$ and $a_A$ are the normalization factors. Values of these parameters found from the fit are shown in 
Table \ref{table}. It is remarkable that the found exponents $b_A$ are close to the critical exponent 
$\beta_{Ising}=0.3265\pm0.0001$ of the 3-dimensional Ising model \cite{Campostrini} and to the 
critical exponent $\beta_{liquids}=0.335\pm0.015$ of simple liquids \cite{Huang}. \vspace*{-.4cm}

\begin{table}[h]
\vspace*{-.2cm}
\centering
\caption{The fit parameters according to Eq. (\ref{VI}) and Eq. (\ref{VII}) }
\vspace*{.1cm}
\label{table}   
{\small    
\begin{tabular}{cccccccc}
\hline
                &          &   \multicolumn{3}{c}{surface tension}  & \multicolumn{3}{c}{average maximal cluster}  \\  \cline{3-8}
 $L_{cut}$ & Type  &     $d_A$    &      $B_A$   & $\chi^2/dof$ &     $a_A$   &       $b_A$    & $\chi^2 / dof$  \\ \hline
       0.1    &   Cl     & 0.485(14)   & 0.2920(12) &    1.43 /4      &  3056(246) & 0.2964(284) &   16.32 /4         \\ \hline 
       0.1    &  aCl    & 2.059(28)   & 0.4129(77) &    1.68/4        &  2129(160) & 0.3315(269) &    8.94/4          \\ \hline
       0.2    &   Cl     & 0.2796(118)& 0.2891(16) &    1.11/4       &  4953(443) & 0.3359(289) &   12.3/3            \\ \hline 
       0.2    &  aCl    & 1.344(33)   & 0.4483(21)  &    0.66/2       & 2462(88)    & 0.3750(129) &    2.068/4         \\ \hline
\end{tabular}}
\vspace*{-.7cm}
\end{table}

\section{Conclusions}

In this contribution  we present a novel approach to study the deconfinement PT  in the SU(2) GD in terms 
of the geometrical clusters composed of the Polyakov loops of the same sign. We justify the separation of
(anti)clusters into ``liquid" droplet and ``gas" of smaller fragments and investigate their physical properties. 
We also explain the deconfinement PT as a special kind of the liquid-gas transition between two types of liquid whose 
behavior is drastically  different in the region of broken global Z(2) symmetry. Above  PT the cluster liquid droplet 
evaporates, whereas the anticluster liquid droplet condensates  the accompanying 
gas of anticlusters. A successful application of the LDF to the description of the size distributions of all gaseous 
(anti)clusters excluding the monomers is the main  result of this study. This approach allowed us to determine the 
$\beta$-dependences  of the reduced chemical potential and the reduced surface tension coefficient. While in 
the phase of unbroken global Z(2) symmetry  these quantities are identical for the fragments of  both kinds,  their behavior is drastically 
different in the deconfined phase. Another  important finding of this study is a high precision determination of 
the Fisher topological constant $\tau=1.806\pm0.008$ which has the same value   both for  clusters and for anticlusters. 
This result is in line with the exactly solvable model of the nuclear liquid-gas PT \cite{Sagun} and disproves the 
FDM prediction that $\tau > 2$ \cite{Fisher67,Fisher69}. We argue   that  the reduced surface tension coefficient 
and the mean size of the largest (anti)cluster can be used as  the new order parameters of  deconfinement  PT in 
SU(2) GD. \vspace*{-.4cm}

\ack

The authors thank A. V. Taranenko
for the fruitful discussions, valuable comments and  kind help in the ICPPA 2016 participation. This  work was supported 
in part by the National Academy of Sciences of Ukraine and by the NAS of Ukraine grant of GRID simulations for high 
energy physics.
\vspace*{-.4cm}

\section*{References}


\end{document}